\documentclass[prb,showpacs,superscriptaddress]{revtex4}
\usepackage{graphicx}
\usepackage{amsmath}
\usepackage{amssymb}
\usepackage{color}

\begin{document}

\title{Edge states, mass and spin gaps, and quantum Hall effect in graphene}

\date{\today}

\preprint{UWO-TH-08/11}

\author{V.P. Gusynin}
\affiliation{Bogolyubov Institute for Theoretical Physics, 03680, Kiev, Ukraine}

\author{V.A. Miransky}
\altaffiliation[On leave from ]{Bogolyubov Institute for Theoretical Physics,
03680, Kiev, Ukraine}
\affiliation{ Department of Applied Mathematics, University of Western Ontario,
London, Ontario, Canada N6A 5B7}

\author{S.G. Sharapov}
\affiliation{Department of Physics, Western Illinois University, Macomb, Illinois 61455, USA}

\author{I.A.~Shovkovy}
\altaffiliation[On leave from ]{Bogolyubov Institute for Theoretical Physics,
03680, Kiev, Ukraine}
\affiliation{Department of Physics, Western Illinois University, Macomb, Illinois 61455, USA}

\date{\today}

\begin{abstract}
Motivated by recent experiments and a theoretical analysis of the 
gap equation for the propagator of Dirac quasiparticles, we assume 
that the physics underlying the recently observed removal of sublattice 
and spin degeneracies in graphene in a strong magnetic field is 
connected with the generation of both Dirac masses and spin gaps.
The consequences of such a scenario for the existence of the gapless 
edge states with zigzag and armchair boundary conditions are discussed. 
In the case of graphene on a half-plane with a zigzag edge, there are 
gapless edge states in the spectrum only when the spin gap dominates 
over the mass gap. In the case of an armchair edge, however, the 
existence of the gapless edge states depends on the specific type 
of mass gaps.
\end{abstract}


%73.20.-r – Electron states at surfaces and interfaces
%73.43.-f   Quantum Hall effects
%73.43.Cd   Theory and modeling
%71.70.Di       Landau levels
%81.05.Uw   Carbon, diamond, graphite

\maketitle

\section{Introduction}
\label{sec:intro}

A graphite monolayer, or graphene, has become a new exciting topic in
physics of two-dimensional electronic systems.\cite{Novoselov2004Science,rev1,rev2}
A qualitatively new feature of graphene is that its low-energy quasiparticles
are described by a relativistic $2+1$-dimensional Dirac 
theory.\cite{Semenoff1984PRL,DiVincenzo1984PRB,Haldane1988PRL} 
The spinor structure of the corresponding wave functions is a consequence 
of the honeycomb lattice structure of graphene with two carbon atoms per 
unit cell. When a magnetic
field is applied, noninteracting Dirac quasiparticles occupy the Landau
levels (LLs) with the following energies:
\begin{equation}
\label{E_n}
E_n = \pm \sqrt{2 n \hbar v_F^2 |eB|/c}
 \approx \pm  424 \sqrt{n}  \sqrt{B[\mbox{T}]} \, \mbox{K} ,
\end{equation}
with $n =0, 1, 2, \ldots$. Here $B$ is the value of the magnetic field orthogonal 
to the graphene's plane and $v_F \approx 10^{6} \mbox{m/s}$ is the Fermi 
velocity.

Several anomalous properties of graphene are attributed to the presence
of the lowest Landau level (LLL), i.e., the $n=0$ state in spectrum (\ref{E_n}),
whose energy is independent of the field strength. For example, the
anomaly manifests itself as the phase shift of $\pi$
in the quantum magnetic oscillations of the diagonal conductivity.
This phase shift can be theoretically understood by using either
the semiclassical quantization condition for quasiparticles with a linear
dispersion,\cite{Mikitik1999PRL} or a microscopic calculation for both
massless and massive Dirac fermions.\cite{Sharapov2004PRB,Gusynin2005PRB}
In the Hall conductivity, the anomaly results in an unconventional integer
quantum Hall (QH) effect with the plateaus at the filling factors
$\nu = \pm 4(n+1/2)$.\cite{Zheng2002PRB,Gusynin2005PRL,
Gusynin2006PRB,Peres2006PRB} These and other distinct
properties of graphene allow one to unambiguously identify the Dirac
nature of quasiparticles in experiments.\cite{Geim2005Nature,Kim2005Nature}

While many unusual properties of graphene can be explained by using
the framework of a noninteracting Dirac theory, the quasiparticle
interactions are not negligible. In fact, they are responsible for the
appearance of additional QH plateaus with the filling factors
$\nu =0,\, \pm1,\, \pm 4$ that were first reported in
Ref.~\onlinecite{Zhang2006PRL} in the case of sufficiently
strong magnetic fields, $B \gtrsim 20\, \mbox{T}$ (see also
Refs.~\onlinecite{Abanin2007PRL,Kim2007PRL,Giesbers2007PRL,Ong2007}).

Recently, we proposed a dynamical mechanism,\cite{Gusynin2006cat} which is based 
on the phenomenon of the magnetic catalysis,\cite{Gusynin1994PRL} that 
could explain the $\nu=0$ and $\nu = \pm 1$ plateaus in the Hall conductivity 
of graphene.\cite{Zhang2006PRL} The subsequent
experiments\cite{Abanin2007PRL,Kim2007PRL} have revealed several
additional features of the $\nu=0$ and $\nu = \pm 1$ plateaus that seem to
require modifications of the scenario in Ref.~\onlinecite{Gusynin2006cat}.
Among them, the most important is a rather peculiar dissipative nature of 
the diagonal transport at the $\nu=0$ plateau. 
This seems to suggest that the origin of the $\nu=0$ plateau is associated 
with a spin gap rather than a mass gap.\cite{Abanin2007PRL,Abanin2006PRL} This
conclusion is supported by the fact that the activation energy at the $\nu=0$
plateau is vanishing.\cite{Abanin2007PRL,Kim2007PRL} Additionally,
the diagonal transport is suggested to be dominated by gapless edge
states, which should exist when the lowest Landau level is split by a
large spin gap.\cite{Abanin2007PRL,Abanin2006PRL}

Concerning the $\nu =\pm 1$ plateaus, the measurements of the thermal
activation energy $\Delta E (\nu = \pm 1)$ point to its connection with orbital
dynamics. Indeed, the activation energy depends only on the perpendicular
component of the magnetic field.\cite{Zhang2006PRL,Kim2007PRL}
The dynamical nature of the $\nu =\pm 1$ plateaus is also suggested 
by the fact that $\Delta E (\nu = \pm 1)$ is proportional to 
$\sqrt{B}$.\cite{Zhang2006PRL,Kim2007PRL}

Note that, in contrast, the $\nu =\pm 4$ plateaus can be consistently associated with
the Zeeman splitting of the $n=1$ Landau level. The corresponding activation
energy $\Delta E (\nu = \pm 4)$ linearly depends on the total magnetic field and
has the same magnitude as the Zeeman energy,\cite{Zhang2006PRL,Kim2007PRL}
\begin{equation}
\label{Zeeman-usual} E_Z = \frac{g_L}{2} \mu_B B \simeq
0.67\, B[\mbox{T}] \, \mbox{K} ,
\end{equation}
where $\mu_B = e\hbar/(2 m c)$ is the Bohr magneton and $g_L \simeq 2$ is the
Lande factor in graphene.

Theoretically, the $\nu=0$ and $\nu =\pm 1$ plateaus come from lifting the
approximate degeneracy of the four sublevels at LLL. The degeneracy is a
consequence of the ``flavor" $U(4)$ symmetry of the low-energy continuum 
description of graphene in the absence of a Zeeman interaction. This symmetry
operates in the space of the sublattice-valley and spin degrees of freedom.
If it is accepted that the $\nu=0$ plateau is due to a spin gap, then the $\nu =\pm 1$
plateaus should result from breaking the sublattice-valley symmetry. This seems
to be in agreement with the observations in Ref.~\onlinecite{Kim2007PRL}.

There are essentially two approaches that consider various possibilities of
breaking the approximate $U(4)$ symmetry of graphene (for a brief review, 
see Ref.~\onlinecite{Yang2007SSC}).

\begin{itemize}
\item[(i)] 
The quantum Hall ferromagnetism (QHF) scenario\cite{Nomura2006PRL,
Goerbig2006,Alicea2006PRB} {which is connected with the theory of
exchange-driven spin-splitting of Landau levels in Ref.~\onlinecite{Fogler1995PRB}.}
It exploits an analogy between the four-fold degeneracy of LLs in graphene, 
which is associated with the $U(4)$ symmetry, and the $SU(4)$ ferromagnetism 
previously studied in the bilayer quantum Hall systems.\cite{foot1} In this 
scenario the QH plateaus with all integer values of the filling factor $\nu$ 
occur in sufficiently clean samples. The QHF order parameters are described 
by the densities of conserved charges connected with the diagonal generators 
of the $SU(4) \subset U(4)$ symmetry group.

\item[(ii)]
The magnetic catalysis (MC) scenario\cite{Gusynin2006cat,Fuchs2006,
Herbut2006,Ezawa2006} that uses the idea of a spontaneous symmetry 
breaking due to the exciton (chiral) condensation.\cite{Gusynin1994PRL,
Khveshchenko2001PRL,Gorbar2002PRB,Gorbar2003PLA} Such a 
condensation produces a nonzero Dirac mass term in the low-energy theory of 
graphene. (Note that originally the magnetic catalysis scenario in graphene was 
motivated by the early experiments in highly oriented pyrolytic graphite.\cite{kopel1})
\end{itemize}

\begin{figure}
\includegraphics[width=.48\textwidth]{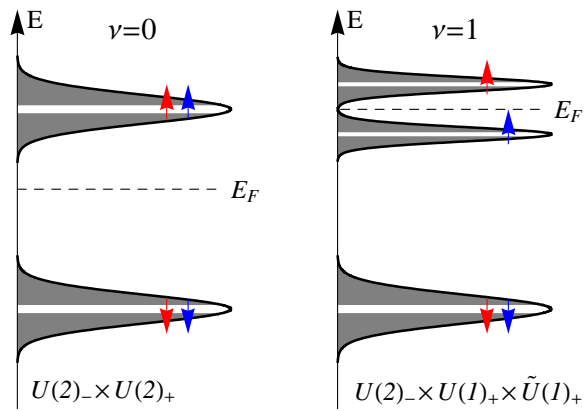}
\caption{Illustration of the lowest Landau level splitting needed
to explain $\nu=0$ and $\nu=1$ plateaus in QHE in graphene.}
\label{fig-illustration}
\end{figure}

As emphasized in Ref.~\onlinecite{Gusynin2006cat}, the plateau
$\nu =0$ could appear due to either an enhanced spin gap or a mass
term. An enhanced spin gap breaks the approximate $U(4)$ symmetry 
down to the $U(2)_{-}\times U(2)_{+}$ subgroup which operates in the 
sublattice-valley space and does not mix spin-up ($s=+$) and spin-down 
($s=-$) states. A nonzero Dirac mass term breaks the symmetry down 
to another $U(2)\times U(2)^\prime$ subgroup, which operates in the spin 
space. Either of them is sufficient to partially lift the four-fold 
degeneracy of the LLL that is needed in the $\nu=0$ QH state. The
structure of the energy sublevels at LLL in the case of a nonzero spin
gap is illustrated in the left panel of Fig.~\ref{fig-illustration}.

%%%%%%%%%%%%%%%%%%%%%%%%%%%%%%%%%%
%%%%%%%%%%%%%%%%%%%%%%%%%%%%%%%%%%
%%%%%%%%%%%%%%%%%%%%%%%%%%%%%%%%%%
%%%%%%%%%%%%%%%%%%%%%%%%%%%%%%%%%%
%%%%%%%%%%%%%%%%%%%%%%%%%%%%%%%%%%
%%%%%%%%%%%%%%%%%%%%%%%%%%%%%%%%%%

In order to explain the $\nu = \pm 1$ QH plateaus, two different order
parameters are required. (Note that the choice of two order parameters 
with given symmetry properties is not unique.\cite{GGM2007}) This 
should already be evident from the symmetry arguments alone. For 
example, the simplest possible structure of the energy sublevels for 
the $\nu = +1$ state is shown in the right panel of Fig.~\ref{fig-illustration}. 
The corresponding splitting is possible only if the $U(2)_{-}\times U(2)_{+}$  
symmetry is further reduced, e.g.,  at least down to the $U(2)_{-} \times 
U(1)_{+}\times \tilde{U}(1)_{+}$ subgroup. However, this would not be 
possible without having an additional order parameter that breaks the 
sublattice-valley symmetry, which is described by the simple Lie group 
$SU(2)_{+}\subset U(2)_{+}$.

An approach that combines both QHF and MC mechanisms in a unifying
scheme was recently proposed in Ref.~\onlinecite{GGM2007}. By making
use of a multi-parameter variational ansatz for the quasiparticle propagator,
it was found that QHF ($\mu_s$  and $\tilde\mu_s$) and MC ($\Delta_s$
and $\tilde\Delta_s$) order parameters necessarily coexist. In terms
of symmetry, the order parameters of the first type, i.e., $\mu_s$
and $\Delta_s$ with nonequal values for $s=\pm$, break the $U(4)$
symmetry down to $U(2)_{-}\times U(2)_{+}$ just like the Zeeman
term. The order parameters of the other type, $\tilde\mu_s$ and 
$\tilde\Delta_s$, are triplets with respect to the $SU(2)_{s}$ group, which 
is the largest non-Abelian subgroup of the $U(2)_{s}$. Thus, when either 
$\tilde\mu_s$ or $\tilde\Delta_s$ has a nonzero vacuum expectation
value, the symmetry $SU(2)_{s}$ is further broken down to $U(1)_{s}$.

The motivation for the present work is to address the question of
compatibility of the microscopic dynamics described in
Ref.~\onlinecite{GGM2007} with the gapless edge states, which are 
apparently needed in the $\nu=0$ state.\cite{Abanin2007PRL,Kim2007PRL}
Our main results are as follows. In the case of graphene on a half-plane
with a zigzag edge, there are gapless edge states in the spectrum only 
when the spin gap dominates over the mass gap. In the case of an armchair 
edge, however, the existence of the gapless edge states depends on 
the specific types of mass gaps. As will be discussed below, these 
results could have important consequences for understanding dynamics 
in the QH effect in graphene.

The paper is organized as follows. In Sec.~\ref{sec:model} we present
a model Lagrangian that captures the most general dynamical situation
with QHF and MC order parameters, as proposed in Ref.~\onlinecite{GGM2007}.
The spectrum of the corresponding Dirac equation in an external magnetic
field is analyzed in Sec.~\ref{sec:Dirac-magnetic}. The edge states for zigzag
and armchair edges are considered in Secs.~\ref{sec:zigzag} and \ref{sec:armchair},
respectively. The main results of the paper are discussed in Sec.~\ref{sec:concl}.

\section{Model with dynamical gaps}
\label{sec:model}

The low-energy quasiparticle excitations in graphene are described
in terms of a four-component Dirac spinor $\Psi_{s}^T= \left(
\psi_{K_{+}As},\psi_{K_{+}Bs},\psi_{K_{-} Bs}, \psi_{K_{-} As}\right)$.
The spinor (with a given spin index $s=\pm$) combines the Bloch states
on the two different sublattices ($A$ and $B$) of the hexagonal graphene
lattice and with the momenta near the two inequivalent Dirac points ($K_+$ 
and $K_-$) of the two-dimensional Brillouin zone. The quadratic part 
of low-energy Lagrangian density for quasiparticles can be written in 
a relativistic form as
\begin{equation}
\label{Lagrangian}
\mathcal{L} = \sum_{s = \pm} \hbar \,
\bar{\Psi}_{s} (t, \mathbf{r}) \left(
i \gamma^0 \partial_t
+ i v_F \gamma^1 D_x
+ i v_F \gamma^2 D_y \right)
\Psi_{s}(t, \mathbf{r})
+\mathcal{L}_{\rm mass}
+ \sum_{s = \pm} \left(\mu_s \rho_s+\tilde{\mu}_s \tilde{\rho}_s\right),
\end{equation}
where $\bar{\Psi}_{s} = \Psi_{s}^\dagger \gamma^0$ is the Dirac 
conjugated spinor and the operators $\rho_s\equiv \bar{\Psi}_{s} \gamma^0 \Psi_{s}$ 
and $ \tilde{\rho}_s\equiv\bar{\Psi}_{s} \gamma^0 \gamma^5\Psi_{s}$ are 
densities of conserved charges connected with the chemical potentials 
$\mu_s$ and $\tilde{\mu}_s$ ($s=\pm$), respectively. Notice that here 
the Fermi velocity $v_F\approx c/300$ plays the role of the speed of light. 
The orbital effect of a perpendicular magnetic field 
$\mathbf{B} = \nabla \times \mathbf{A}$ is included via the covariant 
derivative $D_i=\partial_i+(ie/\hbar c)A_i$, where $i=x,y$ and $-e<0$ is 
the electron charge. Here, we assume that the vector potential is taken
in the Landau gauge: $A_x = -B y$ and $A_y = 0$, where $B$ is
the magnitude of the magnetic field. The mass term $\mathcal{L}_{\rm mass}$ 
is defined below.

The $4\times4$ matrices $\gamma^\nu$ furnish a reducible representation
of the Dirac algebra. Here, we use the following representation:
\begin{equation}
\label{chiral-gamma}
\gamma^0={\tilde \tau}_1 \otimes \tau_0,\qquad
\gamma^i=-i {\tilde \tau}_2 \otimes \tau_i,
\end{equation}
where the Pauli matrices $\tilde{\tau}_i$ and $\tau_i$ (as well as the $2\times2$
unit matrices $\tilde{\tau}_0$ and $\tau_0$) act on the valley ($K_+,\, K_-$) and
the sublattice ($A,\, B$) indices, respectively. This representation is derived
from a tight-binding model for graphene.\cite{Gusynin2007review} It is
particularly convenient for our purposes in this study because it provides
a simple form of the boundary conditions at zigzag and armchair
edges. As follows from definition (\ref{chiral-gamma}), the $\gamma$-matrices
satisfy the usual anticommutation relations
$\left\{\gamma^\mu,\gamma^\nu\right\}=2g^{\mu\nu}$, where
$g^{\mu\nu}=\mbox{diag}(1,-1,-1,-1)$. Since the matrix
$\gamma^5\equiv  i \gamma^0 \gamma^1 \gamma^2 \gamma^3$
is diagonal,
\begin{equation}
\gamma^5  = \left(
          \begin{array}{cc}
            I_2 & 0 \\
            0 & -I_2 \\
          \end{array}
        \right),
\end{equation}
this representation is conventionally called chiral. Note that the chirality here
is identified with the valley index ($K_+$ or $K_-$).\cite{Gusynin2007review}

The general expression for the mass term $\mathcal{L}_{\rm mass}$ in
the Lagrangian density may include singlet ($\Delta_s$) as well as triplet
($\tilde{\Delta}_s$) contributions with respect to the valley symmetry group
$SU(2)_{s}$. The appearance of the mass term can be attributed, for
example, to the MC mechanism. In the representation used here, its
explicit form reads\cite{Gusynin2008}
\begin{equation}
\mathcal{L}_{\rm mass}= \sum_{s = \pm}
\bar{\Psi}_{s} (t, \mathbf{r}) \left(
 \Delta_s\gamma^3\gamma^5
-\tilde{\Delta}_s\gamma^3
\right)\Psi_{s}(t, \mathbf{r}).
\label{L-mass}
\end{equation}
Under the time reversal symmetry, the operators associated with the mass
parameters $\Delta_s$ and
$\tilde{\Delta}_s$ are odd and even, respectively. Concerning the triplet
mass term $\tilde{\Delta}_s \bar{\Psi}_{s} \gamma^3\Psi_{s}$, it can also
be written in other equivalent forms, e.g., as 
$\tilde{\Delta}_s\bar{\Psi}_{s} i\gamma^5\Psi_{s}$ or
$\tilde{\Delta}_s\bar{\Psi}_{s} \Psi_{s}$.\cite{foot2}
The latter, in particular, is the usual Dirac mass term.
All of these representations are equivalent because they are related
by the transformations of the $SU(2)_{s}$ symmetry group. For our purposes,
however, it is most convenient to use the form in Eq.~(\ref{L-mass}) which,
as we shall see below, has a simple interpretation in the tight binding model.

In Lagrangian density [Eq.~(\ref{Lagrangian})], the chemical potentials 
$\mu_s$ and $\tilde{\mu}_s$ allow us to accommodate the QHF order 
parameters in the dynamical model of Ref.~\onlinecite{GGM2007}.
Regarding the transformation properties of $\mu_s$ and $\tilde{\mu}_s$
under the flavor symmetry, they are similar to those of $\Delta_s$ and
$\tilde{\Delta}_s$, respectively.

Before proceeding with further analysis, it is instructive to address the
physics interpretation of the mass parameters and chemical potentials in
more detail. To this end, let us write down the explicit expressions for the
corresponding operators in the Lagrangian density in terms of separate
Bloch components of the spinors as follows:
\begin{eqnarray}
\label{triplet_mass}
\tilde{\Delta}_s:&\quad&
{\bar{\Psi}_s \gamma^3 \Psi_s} =
  \psi_{K_{+} As}^\dagger\psi_{K_{+}As}
+ \psi_{K_{-} As}^\dagger \psi_{K_{-}As}
- \psi_{K_{+}Bs}^\dagger \psi_{K_{+}Bs}
- \psi_{K_{-}Bs}^\dagger \psi_{K_{-} Bs}, \\
\label{singlet_mass}
\Delta_s: &\quad&
{\bar{\Psi}_s \gamma^3 \gamma^5 \Psi_s} =
  \psi_{K_{+} As}^\dagger\psi_{K_{+}As}
- \psi_{K_{-} As}^\dagger \psi_{K_{-}As}
- \psi_{K_{+}Bs}^\dagger \psi_{K_{+}Bs}
+ \psi_{K_{-}Bs}^\dagger \psi_{K_{-} Bs},\\
\label{triplet_mu}
\tilde{\mu}_s:&\quad&
{\bar{\Psi}_s \gamma^0 \gamma^5 \Psi_s} =
  \psi_{K_{+} As}^\dagger\psi_{K_{+}As}
- \psi_{K_{-} As}^\dagger \psi_{K_{-}As}
+ \psi_{K_{+}Bs}^\dagger \psi_{K_{+}Bs}
- \psi_{K_{-}Bs}^\dagger \psi_{K_{-} Bs},\\
\label{singlet_mu}
\mu_s:&\quad&
{\bar{\Psi}_s \gamma^0 \Psi_s} =
  \psi_{K_{+} As}^\dagger\psi_{K_{+}As}
+ \psi_{K_{-} As}^\dagger \psi_{K_{-}As}
+ \psi_{K_{+}Bs}^\dagger \psi_{K_{+}Bs}
+ \psi_{K_{-}Bs}^\dagger \psi_{K_{-} Bs}.
\end{eqnarray}
Here the operators on the right hand side are linear combinations of the
electron densities at specified valleys  ($K_+$ or $K_-$) and sublattices
($A$ or $B$). These operators enter into the Lagrangian density together
with the parameters $\Delta_s$, $\tilde{\Delta}_s$, $\mu_s$, and $\tilde{\mu}_s$,
which play the role of Lagrange multipliers. Therefore, the values of  the
masses and chemical potentials control the relative concentrations of
electrons at different valleys and sublattices. They are determined from
the gap equations for Dirac quasiparticles.\cite{GGM2007}

As seen from Eq.~(\ref{triplet_mass}), the triplet Dirac mass $\tilde{\Delta}_s$
is related to the density imbalance between the $A$ and $B$ sublattices.
Its spontaneous generation leads to a state with a charge density
wave.\cite{Gusynin2006cat,Khveshchenko2001PRL,Herbut2006,
Fuchs2006,Ezawa2006} If the values of the dynamical masses are 
nonequal for different spins $s=\pm$, an admixture of an antiferromagnetic 
wave develops in the ground state.\cite{Herbut2006} 
Recent scanning tunneling spectroscopy revealed a mass 
gap near the Dirac point in a single layer graphene sample suspended 
above a graphite substrate.\cite{Andrei} The gap could be interpreted 
as a Dirac mass gap induced by a substrate perturbation that breaks 
the sublattice symmetry.\cite{Giovannetti2007} In the case of epitaxial 
graphene grown on SiC, the presence of a nonzero Dirac gap is strongly 
supported by the angle resolved photoemission spectroscopy
measurements.\cite{Lanzara}

The value of the singlet Dirac mass $\Delta_s$ [see Eq.~(\ref{singlet_mass})]
controls a mixed density imbalance at the two valleys and the two sublattices. 
Similarly, the chemical potential $\tilde{\mu}_s$ is connected with the density 
imbalance between the two valleys and, at last, $\mu_s$ is the usual chemical 
potential related to the total density of electrons with a given spin.

In general, in a finite geometry sample, the magnitude of the exchange and 
Hartree interactions, which determine the values of the parameters $\Delta_s$, 
$\tilde{\Delta}_s$, $\mu_s$, and $\tilde{\mu}_s$, is likely to vary with the 
distance from the edges and should be calculated in a self-consistent way. 
The present study of the edge states is done by assuming uniform gaps and 
uniform chemical potentials. We believe, however, that such an idealized 
treatment should be sufficient to capture the main  qualitative (although not 
quantitative) features of the edge states.

\section{Dirac equation in an external magnetic field}
\label{sec:Dirac-magnetic}

In this section, we study the spectrum of the low-energy quasiparticles
in the model of graphene with the most general set of parameters
$\Delta_s$, $\tilde{\Delta}_s$, $\mu_s$, and $\tilde{\mu}_s$. The
corresponding Dirac equation takes the following form:
\begin{equation}
\label{Dirac-gamma3} \left[ i \gamma^0 \hbar
\partial_t  + i \hbar v_F \gamma^1 D_x + i \hbar v_F \gamma^2 D_y
+ \mu\gamma^{0} +\tilde{\mu}\gamma^{0} \gamma^{5}
+ \Delta\gamma^{3} \gamma^{5}-\tilde{\Delta}\gamma^{3}\right]
\Psi(t, \mathbf{r})=0.
\end{equation}
For brevity of notation, the spin index is omitted here and below. For the
energy eigenvalue solutions $\Psi(t, \mathbf{r})=e^{-iEt/\hbar}\Psi(
\mathbf{r})$, the equation reduces to
\begin{equation}
\label{Dirac-E}
\left[\hbar
v_{F}\left(-\alpha_{1}iD_{x}-\alpha_{2}iD_{y}\right)
-\mu-\tilde{\mu} \gamma^{5}- i \Delta\gamma^{1} \gamma^{2}
+\tilde{\Delta}\alpha_{3}\right]
\Psi( \mathbf{r})=E\Psi( \mathbf{r}), \end{equation}
where the $\alpha$-matrices are
\begin{equation}
\alpha_{i}=\gamma^{0}\gamma^{i}= \left(
\begin{array}{cc}
    \sigma_{i} &  0 \\
    0 &-\sigma_{i}
\end{array}\right).
\end{equation}
By using the representation for the $\gamma$-matrices in Eq.~(\ref{chiral-gamma}),
we can rewrite the Dirac equation in the components as follows:
\begin{eqnarray}
\left(
\begin{array}{cc}
    -\mu^{(+)}-\Delta^{(-)} & - \hbar v_{F}\left(iD_{x}+D_{y}\right) \\
    -\hbar v_{F}\left(iD_{x}-D_{y}\right) & -\mu^{(+)}+\Delta^{(-)}
\end{array}\right)\left(\begin{array}{c}\psi_{AK_{+}}\\
\psi_{BK_{+}}\end{array}\right)&=&E\left(\begin{array}{c}\psi_{AK_{+}}\\
\psi_{BK_{+}}\end{array}\right),
\label{eq:Kplus}\\
\left( \begin{array}{cc}
    -\mu^{(-)}-\Delta^{(+)}  &  \hbar v_{F}\left(iD_{x}+D_{y}\right) \\
    \hbar v_{F}\left(iD_{x}-D_{y}\right) &-\mu^{(-)}+\Delta^{(+)}
\end{array}\right)\left(\begin{array}{c}\psi_{BK_{-}}\\
\psi_{AK_{-}}\end{array}\right)&=&E\left(\begin{array}{c}\psi_{BK_{-}}\\
\psi_{AK_{-}}\end{array}\right).
\label{eq:Kminus}
\end{eqnarray}
Here, we introduced the shorthand notation: $\mu^{(\pm)}\equiv \mu\pm\tilde{\mu}$
and $\Delta^{(\pm)}\equiv \Delta\pm\tilde{\Delta}$.
As we can see, the equations for different valleys decouple. This is a very
useful property that considerably simplifies the analysis. In each of the
two decoupled sets of equations, we can express the $B$-components
in terms of the $A$-components of the spinors,
\begin{eqnarray}
\psi_{BK_{+}} &=& -\frac{\hbar v_{F}\left(iD_{x}-D_{y}\right)}{E+\mu^{(+)}-\Delta^{(-)}}
\psi_{AK_{+}},\\
\psi_{BK_{-}} &=& \frac{\hbar v_{F}\left(iD_{x}+D_{y}\right)}{E+\mu^{(-)}+\Delta^{(+)}}\psi_{AK_{-}}.
\end{eqnarray}
Then, at $K_{+}$ and $K_{-}$ valleys, the two-component spinors can be
written in the following form:
\begin{eqnarray}
\label{eliminated}
\psi_{K_{+}} &=&A_{1}\left(\begin{array}{c} \psi_{AK_{+}}\\
-\frac{\hbar v_{F}\left(iD_{x}-D_{y}\right)}{E+\mu^{(+)}-\Delta^{(-)}}
\psi_{AK_{+}}\end{array}\right),\\
\psi_{K_{-}}&=&A_{2}\left(\begin{array}{c}
\frac{\hbar v_{F}\left(iD_{x}+D_{y}\right)}{E+\mu^{(-)}+\Delta^{(+)}}\psi_{AK_{-}}\\
\psi_{AK_{-}}
\end{array}\right).
\end{eqnarray}
Here, the constants $A_{1,2}$ are determined by the normalization
conditions,
\begin{equation}
\label{normalization} \int d^2 r
\psi^\dagger_{K_{\pm}}(\mathbf{r},k,n)\psi_{K_{\pm}}(\mathbf{r},k^{\prime},n^{\prime})
=\delta_{n,n^{\prime}}\delta(k-k^{\prime}),
\end{equation}
where $k,k^{\prime}$ and $n,n^{\prime}$ are the quantum numbers (e.g.,
the wave vector along the $x$ or $y$ direction and the Landau level index)
 that characterize the eigenstates of Dirac quasiparticles in the magnetic
field.

As follows from Eqs.~(\ref{eq:Kplus}) and (\ref{eq:Kminus}), the
$A$-components of the spinors satisfy the following second order
differential equations:
\begin{equation}
\label{differential-common}
\begin{split}
&\left(-l^2 D_{x}^{2}-l^2 D_{y}^{2}+1\right)\psi_{AK_{+}}=2 \lambda_{+}\psi_{AK_{+}},\\
&\left(-l^2D_{x}^{2}-l^2D_{y}^{2}-1\right)\psi_{AK_{-}}=2\lambda_{-}\psi_{AK_{-}}.
\end{split}
\end{equation}
Here we introduced the two dimensionless parameters $\lambda_{\pm} \equiv
\left[\left(E+\mu^{(\pm)}\right)^{2}-\left(\Delta^{(\mp)}\right)^{2}\right]/\epsilon_0^{2}$,
the Landau energy scale $\epsilon_0\equiv \sqrt{2\hbar v_F^2 |eB|/c}$, and the
magnetic length $l \equiv \sqrt{\hbar c/|eB|}$.

In the Landau gauge $(A_{x},A_{y})=(-By,0)$, the differential equations in
Eq.~(\ref{differential-common}) do not explicitly depend on the $x$-coordinate,
and therefore, the wave functions are plane waves in the $x$-direction,
\begin{equation}
\label{x-zigzag}
\begin{split}
\psi_{AK_{+}}(\mathbf{r},k)&=\frac{1}{\sqrt{2\pi
l}}e^{ikx}u_{+}(y,k),\qquad \psi_{BK_{+}}=\frac{1}{\sqrt{2\pi
l}}e^{ikx}v_{+}(y,k), \\
\psi_{AK_{-}}(\mathbf{r},k)&=\frac{1}{\sqrt{2\pi
l}}e^{ikx}u_{-}(y,k), \qquad \psi_{BK_{-}}=\frac{1}{\sqrt{2\pi
l}}e^{ikx}v_{-}(y,k),
\end{split}
\end{equation}
where the functions $u_{\pm}(y,k)$ depend only on a single combination
of the variables, $\xi=y/l-kl$, and satisfy the following equations:
\begin{equation}
\label{differential-zigzag}
\left(\partial^{2}_{\xi}-\xi^{2}\mp1+2\lambda_{\pm} \right)u_{\pm}(\xi)=0.
\end{equation}
In accordance with Eq.~(\ref{eliminated}), the eliminated components
$v_{\pm}(y,k)\equiv v_{\pm}(\xi)$ are given by
\begin{equation}
\label{v_pm-zigzag}
v_{\pm}(\xi)=\frac{\epsilon_0 \left(\partial_{\xi}\mp\xi\right)u_{\pm}(\xi)}
{\sqrt{2}(E+\mu^{(\pm)}\mp\Delta^{(\mp)})}.
\end{equation}
In an infinite system without boundaries, normalizable solutions to 
Eq.~(\ref{differential-zigzag}) are expressed in terms of the Hermite polynomials,
$u(\xi),\, v(\xi) \propto e^{-\xi^2/2}H_n(\xi)$, provided the parameters
$\lambda_{\pm}$ take nonnegative integer values, i.e.,
\begin{equation}
\label{lambda_n}
\lambda_{\pm}=n, \quad \mbox{where} \quad n=0,1,2\dots.
\end{equation}
Note that the value of the energy $E=-\mu^{(+)}+\Delta^{(-)}$ corresponds
to a normalizable LLL state in the $K_{+}$ valley. For such a state, the
apparent singularity in the $v_{+}(\xi)$ component of the wave function 
[see Eq.~(\ref{v_pm-zigzag})] is removed by a proper redefinition
of the normalization constant. The same is not true, however, for the
value of the energy $E=-\mu^{(-)}-\Delta^{(+)}$ in the $K_{-}$ valley.
In fact, a direct analysis shows that the only $n=0$
state in the $K_{+}$ valley has energy $E=-\mu^{(+)}+\Delta^{(-)}$ and
resides solely on the $B$ sublattice, while the only $n=0$ state in the
$K_{-}$ valley has energy $E=-\mu^{(-)}+\Delta^{(+)}$ and resides
solely on the $A$ sublattice.

\section{Edge states for the zigzag edge}
\label{sec:zigzag}

\begin{figure}
\begin{center}
\includegraphics[width=.48\textwidth]{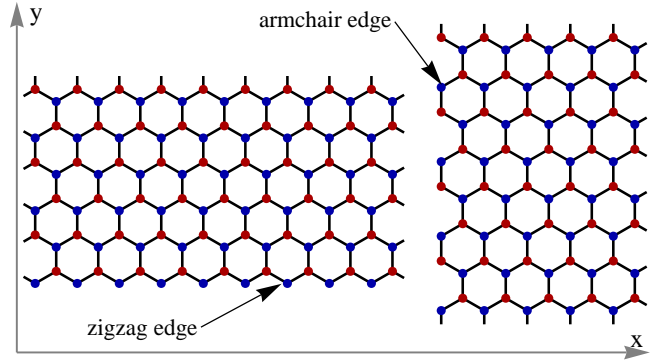}
\caption{Graphene lattice with zigzag and armchair edges.}
\label{illustration}
\end{center}
\end{figure}

There exist many studies of edge states in graphene under various 
conditions.\cite{Peres2006PRB,Abanin2007PRL,Abanin2006PRL,
Gusynin2008,Nakada,Fujita,Falko2004JPCM,KaneMele,Brey2006PRB,
PCNG2006,Peeters} Here we consider a graphene monolayer on the 
half-plane $y>0$ with a zigzag edge parallel to $x$, 
as shown in Fig.~\ref{illustration}.
%%%%%%%%%%%%%%%%%%%%%%%%%%%%%%%%%%%%%%%%%
%%%%%%%%%%%%%  Figure  illustration  %%%%%%%%%%%%%%%%%%
%%%%%%%%%%%%%%%%%%%%%%%%%%%%%%%%%%%%%%%%%
%%%%%%%%%%%%%%%%%%%%%%%%%%%%%%%%%%%%%%%%%
%%%%%%%%%%%%%%%%%%%%%%%%%%%%%%%%%%%%%%%%%
%%%%%%%%%%%%%%%%%%%%%%%%%%%%%%%%%%%%%%%%%
To obtain the energy spectrum we need to supplement the differential equations
for the $u_{\pm}(y,k)$  and $v_{\pm}(y,k)$ functions with suitable boundary conditions.
Such conditions can be derived from the tight-binding model.\cite{Falko2004JPCM,
Brey2006PRB,Abanin2006PRL} For example, for a zigzag edge parallel to the $x$
axis, the wave function on the $A$ atoms should vanish at $y=0$,
\begin{equation}
\label{boundary-zigzag-A}
u_{+}(y=0)=u_{-}(y=0) =0.
\end{equation}
The general solution to Eq.~(\ref{differential-zigzag}) is expressed in
terms of the parabolic cylinder (Weber) functions $U(a,z)$ and
$V(a,z)$,\cite{Abramowitz}
\begin{eqnarray}
\label{uPlus}
&&
u_{+}(\xi)= C_{1} \frac{E+\mu^{(+)}-\Delta^{(-)}}{\epsilon_0 }
U\left(\frac{1-2\lambda_{+}}{2},\sqrt{2}{\xi}\right)
+C_{2} V\left(\frac{1-2\lambda_{+}}{2},\sqrt{2}{\xi}\right),\\
\label{uMinus}
&&
u_{-}(\xi) = C_{3}U\left(-\frac{1+2\lambda_{-}}{2},\sqrt{2}{\xi}\right)
+C_{4} \frac{E+\mu^{(-)}+\Delta^{(+)}}{\epsilon_0}
V\left(-\frac{1+2\lambda_{-}}{2},\sqrt{2}{\xi}\right).
\end{eqnarray}
Here, for convenience of further analysis, the integration
constants $C_{1}$ and $C_{4} $ are introduced together with the
additional factors $\left(E+\mu^{(+)}-\Delta^{(-)}\right)/\epsilon_0$
and $\left(E+\mu^{(-)}+\Delta^{(+)}\right)/\epsilon_0$, respectively.

In an infinite system without edges, the normalizable wave functions
contain only the parabolic cylinder $U(a,z)$-functions, which are 
bound at $z\to\pm\infty$, provided $a=-n-1/2$ and $n$ is a nonnegative
integer. In this case, the following relation is valid:
$U(-n-1/2,z)=2^{-n/2} e^{-z^2/4}H_n(z/\sqrt{2})$,
where $H_n(z)$ are the Hermite polynomials. Therefore, as stated
in Sec.~\ref{sec:Dirac-magnetic}, the spectrum is given by $\lambda_{\pm}=n$
where $n=0,1,2,\ldots$. (A special nature of LLL should be kept in
mind: at $n=0$ there are only two rather than four possible energy
eigenvalues that correspond to normalizable states.)

By using the following recurrent relations for parabolic cylinder 
functions,\cite{Abramowitz}
\begin{eqnarray}
\label{D-relations}
\begin{split}
& \left(\frac{d}{dz}+\frac{z}{2}\right)U(a,z)=-\left(a+\frac12\right)U(a+1,z),\\
& \left(\frac{d}{dz}-\frac{z}{2}\right)U(a,z)=U(a-1,z), \\
& \left(\frac{d}{dz}+\frac{z}{2}\right)V(a,z)=V(a+1,z), \\
& \left(\frac{d}{dz}-\frac{z}{2}\right)V(a,z)=\left(a-\frac12\right)V(a-1,z),
\end{split}
\end{eqnarray}
and Eq.~(\ref{v_pm-zigzag}), we obtain the $v_{\pm}({\xi})$ functions,
\begin{eqnarray}
\label{vPlus}
&&
v_{+}({\xi})=-C_{1}
U\left(-\frac{1+2\lambda_{+}}{2},\sqrt{2}{\xi}\right)
-C_{2}\frac{E+\mu^{(+)}+\Delta^{(-)}}{\epsilon_0 }
V\left(-\frac{1+2\lambda_{+}}{2},\sqrt{2}{\xi}\right),\\
\label{vMinus}
&&
v_{-}({\xi})=C_{3} \frac{E+\mu^{(-)}-\Delta^{(+)}}{\epsilon_0}
U\left(\frac{1-2\lambda_{-}}{2},\sqrt{2}{\xi}\right)
+C_{4}V\left(\frac{1-2\lambda_{-}}{2},\sqrt{2}{\xi}\right).
\end{eqnarray}
On a half-plane, the normalizable wave functions are also given in
terms of only $U(a,z)$-function, which falls off exponentially as
$z\to+\infty$, while the function $V(a,z)$ is growing exponentially
in both directions $z\to\pm\infty$. Therefore, we must take
$C_2=0$ and $C_4=0$. In contrast to the case of an infinite
plane, on a half-plane, there is no restriction for the parameter 
$a$ to be a negative half-integer.

With $C_2=C_4=0$, the zigzag boundary conditions
[(\ref{boundary-zigzag-A})] lead to the following system of equations
\begin{equation}
\label{bc-zigzag-system}
\begin{split}
&C_{1} \left(E+\mu^{(+)}-\Delta^{(-)}\right) D_{\lambda_{+}-1}(-\sqrt{2}kl)=0,\\
&C_{3} D_{\lambda_{-}}(-\sqrt{2}kl)=0.
\end{split}
\end{equation}
Here we introduced another parabolic cylinder function, $D_{\nu}(z)$,\cite{Abramowitz}
which is related to function $U(a,z)$ in a simple way,
\begin{equation}
U(a,z) = D_{-a-1/2}(z).
\label{UD-connection}
\end{equation}
%The $V(a,z)$ function is defined in terms of $D_{\nu}(z)$ as follows:
%\begin{equation}
% V(a,z) = \frac{\Gamma(a+1/2)}{\pi}\left[\sin(\pi a)D_{-a-1/2}(z)+D_{-a-1/2}(-z)\right].
% \end{equation}
There are two types of nontrivial solutions that satisfy the boundary conditions
[(\ref{bc-zigzag-system})]. First, by taking $C_{1}\neq0$ and $C_{3}=0$, we
find that the equation for the eigenvalues is reduced down to 
$E = -\mu^{(+)}+\Delta^{(-)}$ or 
\begin{equation}
\mbox{I.}
\qquad
D_{\lambda_{+}-1}(-kl\sqrt{2})=0 .
\label{eq-z11}
\end{equation}
The solutions of this type have wave functions with a support only in the $K_+$ valley,
\begin{equation}
\begin{split}
 \mbox{I.}
\qquad
& u_{+}(\xi) =C_{1} \frac{E+\mu^{(+)}-\Delta^{(-)}}{\epsilon_0}
D_{\lambda_{+} -1}\left({\sqrt{2}}\xi \right), \\
& v_{+}(\xi)=-C_{1}D_{\lambda_{+} }\left({\sqrt{2}}\xi \right),
\end{split}
\end{equation}
and $u_{-}(\xi)=v_{-}(\xi)=0$.
The other class of solutions is such that $C_{1}=0$ and $C_{3}\neq 0$, and
the energy eigenvalues satisfy the following equation:
\begin{equation}
\mbox{II.}
\qquad
 D_{\lambda_{-}}(-kl\sqrt{2})=0.
\label{eq-z22}
\end{equation}
The wave functions for this type of solutions are nonvanishing only in the $K_-$ valley, i.e.,
\begin{equation}
\begin{split}
\mbox{II.} \qquad
& u_{-}(\xi) = C_{3}D_{\lambda_{-}}\left({\sqrt{2}}\xi \right),\\
& v_{-}(\xi)=C_{3}  \frac{E+\mu^{(-)}-\Delta^{(+)}}{\epsilon_0} D_{\lambda_{-} -1}\left({\sqrt{2}}\xi \right),
\end{split}
\end{equation}
and $u_{+}(\xi)=v_{+}(\xi)=0$.
By making use of the general properties of the parabolic cylinder function
$D_{\nu}(z)$, we can understand some qualitative features of the energy 
spectrum even without solving the equations numerically. To this end, we 
need to know only that, for real $\nu$ and $z$, the function $D_{\nu}(z)$ 
has no real zeros when $\nu$ is negative, and has exactly $\left[\nu+1\right]$ 
real zeros when $\nu$ is nonnegative.\cite{Bateman} Here $\left[\nu+1\right]$ 
denotes the integer part of $\nu+1$. In view of this property, the necessary 
condition for Eq.~(\ref{eq-z11}) to be satisfied is $\lambda_{+}\geq 1$. By 
also including the possibility of the dispersionless mode, which is determined by 
$E = -\mu^{(+)}+\Delta^{(-)}$, we see that the complete spectrum in the $K_{+}$ 
valley (solutions of type I) has the following general structure:
\begin{equation}
\begin{split}
& E_{0}(k) = -\mu^{(+)}+\Delta^{(-)}, \\
& E_{n}(k) = -\mu^{(+)}\pm \sqrt{ \lambda_{+}(kl, n)\epsilon_0^2 +\left(\Delta^{(-)}\right)^2},
\quad\mbox{where}\quad
\lambda_{+}(kl, n)\geq 1,
\end{split}
\label{spectr_K+}
\end{equation}
where $n=1,2,\ldots$ is an index that labels different branches of solutions. By making 
use of the asymptotic behavior of the parabolic cylinder functions, one can show that 
$\lambda_{+}(kl, n)\simeq n$ when $kl\gg 1$. This is expected since large values of 
$kl$ correspond to the states in the bulk, whose wave functions are localized around 
$\xi\simeq 0$ or equivalently $y/l \simeq kl$. (In a system without edges, the index $n$ 
is identified with the usual Landau level index.) 

Similarly, we can constrain the form of the spectrum in the $K_{-}$ valley (solutions of type II). 
The necessary condition for having a real solution to Eq.~(\ref{eq-z22}) is $\lambda_{-}\geq 0$. 
Thus, the energy spectrum in the $K_{-}$ valley has the following general structure:
\begin{equation}
E_{n}(k) = -\mu^{(-)}\pm \sqrt{ \lambda_{-}(kl, n)\epsilon_0^2 +\left(\Delta^{(+)}\right)^2},
\quad\mbox{where}\quad
\lambda_{-}(kl, n)\geq 0,
\label{spectr_K-}
\end{equation}
where $n=0,1,2,\ldots$. Again, one can show that $\lambda_{-}(kl, n)\simeq n$ when 
$kl\gg 1$. 

\begin{figure}
\begin{center}
\includegraphics[width=.48\textwidth]{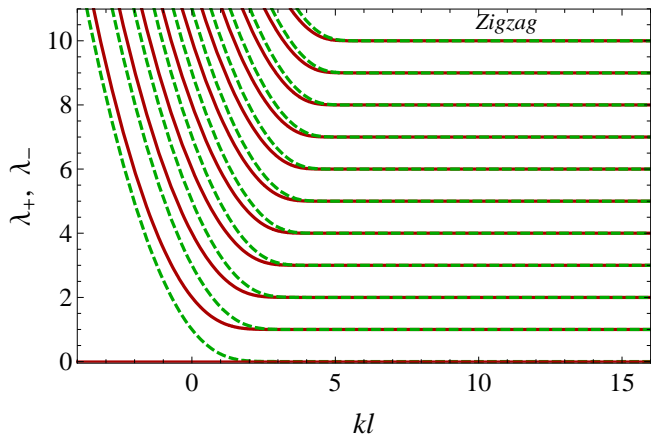}
\caption{The numerical solutions of Eq.~(\ref{eq-z11}) for the dimensionless parameter 
$\lambda_+$ (solid line) and Eq.~(\ref{eq-z22}) for the dimensionless parameter 
$\lambda_-$ (dashed line) in the case of a zigzag boundary. The solid line at
$\lambda_{+} =0$ corresponds only to $E=-\mu^{(+)}+\Delta^{(-)}$ solution.}
\label{fig-zigzag1}
\end{center}
\end{figure}

Our numerical results for $\lambda_{\pm}$ as functions of $k l$ are presented 
in Fig.~\ref{fig-zigzag1}. The solid and dashed lines represent $\lambda_+$ and 
$\lambda_-$, respectively. As expected, there exists an infinite tower of 
solutions that correspond to an infinite tower of Landau levels on a half-plane. 
In Fig.~\ref{fig-zigzag1}, we show only the first 11 solutions.
We also added the constant solution $\lambda_{+}=0$ that,
strictly speaking, represents only the dispersionless mode with the energy
$E = -\mu^{(+)}+\Delta^{(-)}$ [see the first expression in Eq.~(\ref{spectr_K+})].
(Formally, $\lambda_{+}=0$ may also mean that $E = -\mu^{(+)}-\Delta^{(-)}$,
but this is {\em not} an energy eigenvalue.)
%%%%%%%%%%%%%%%%%%%%%%%%%%%%  Figure Zigzag  %%%%%%%%%%%%%%%%%%%%%%%%%%%%
%%%%%%%%%%%%%%%%%%%%%%%%%%%%%%%%%%%%%%%%%%%%%%%%%%%%%%%%%%%%%%%%%%%%%

By analyzing the structure of the spectrum together with the actual 
dependence of  $\lambda_{\pm}$ on the wave vector, we can now
determine when gapless modes exist in the spectrum of graphene 
on a half-plane with a zigzag edge. From Eqs.~(\ref{spectr_K+})
and (\ref{spectr_K-}), we see that the necessary condition to 
have a zero energy state is that at least one of the following 
inequalities is satisfied:
\begin{eqnarray}
K_{+}&\mbox{valley:}&
\qquad
|\mu^{(+)}|\geq \sqrt{\epsilon_0^{2} +\left(\Delta^{(-)}\right)^{2}},
\label{cond-K+}\\
K_{-}&\mbox{valley:}&
\qquad
|\mu^{(-)}|\geq |\Delta^{(+)}|. 
\label{cond-K-}
\end{eqnarray}
From the fact that there exist branches with $\lambda_{+}\simeq 1$ 
and $\lambda_{-}\simeq 0$ at $kl\gg 1$, we see that this is also 
the sufficient condition.

An important point to emphasize here is that nonzero masses do 
not prevent the existence of the gapless edge states when the 
absolute value of $\Delta^{(+)}$ is less than the absolute value of 
$\mu^{(-)}$ at least for one choice of the spin. This is very similar to 
the conditions on a graphene ribbon of finite width,\cite{Gusynin2008} 
except that there are no edge states associated with the second edge
in the present work. Our results generalize the findings of previous 
studies on a half-plane,\cite{Abanin2007PRL,Abanin2006PRL} where 
only the case with a single nonzero order parameter (either mass 
or spin gap) was considered.

\begin{figure}
\begin{center}
\includegraphics[width=.45\textwidth]{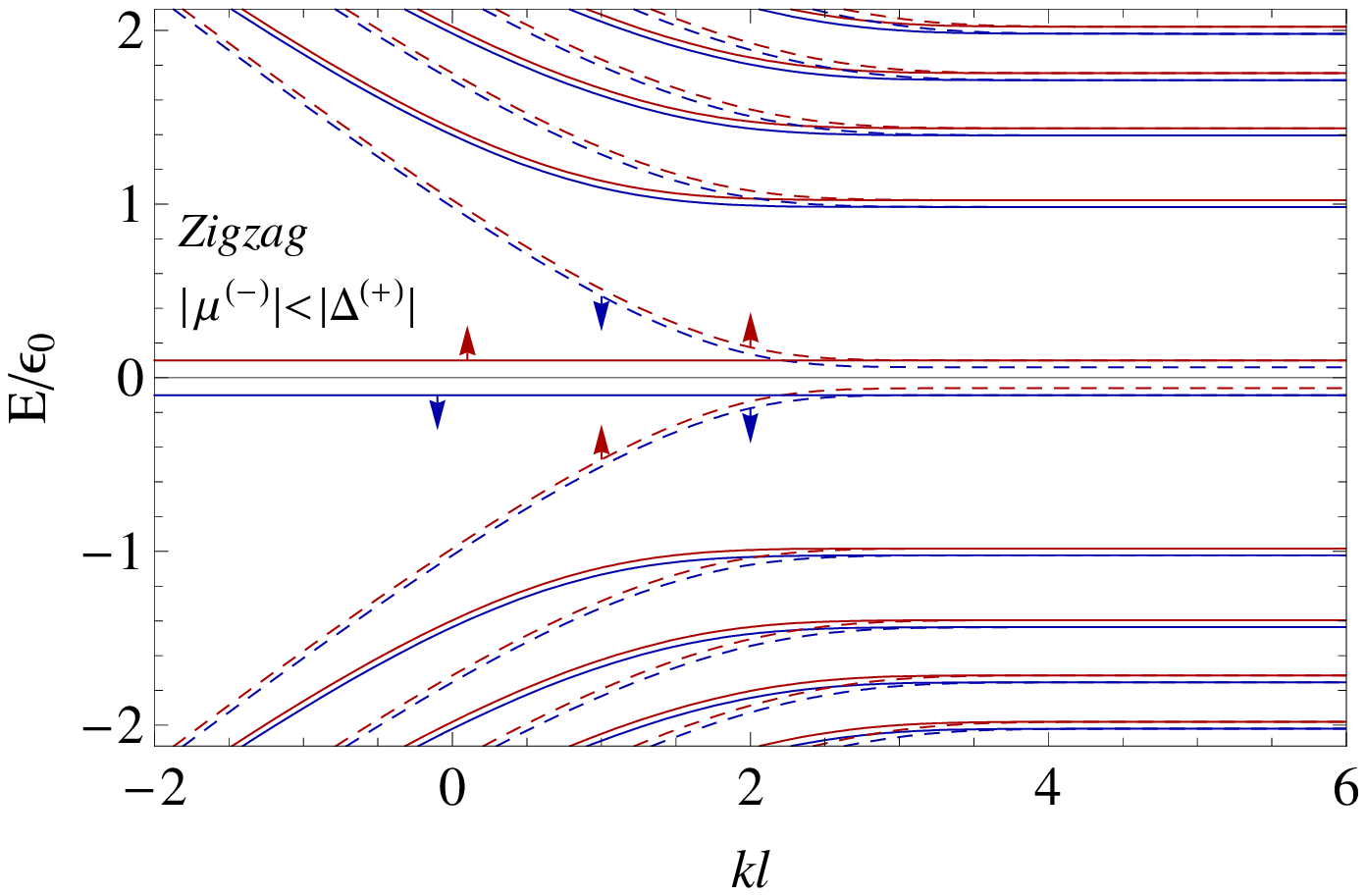}
\hspace{.04\textwidth}
\includegraphics[width=.45\textwidth]{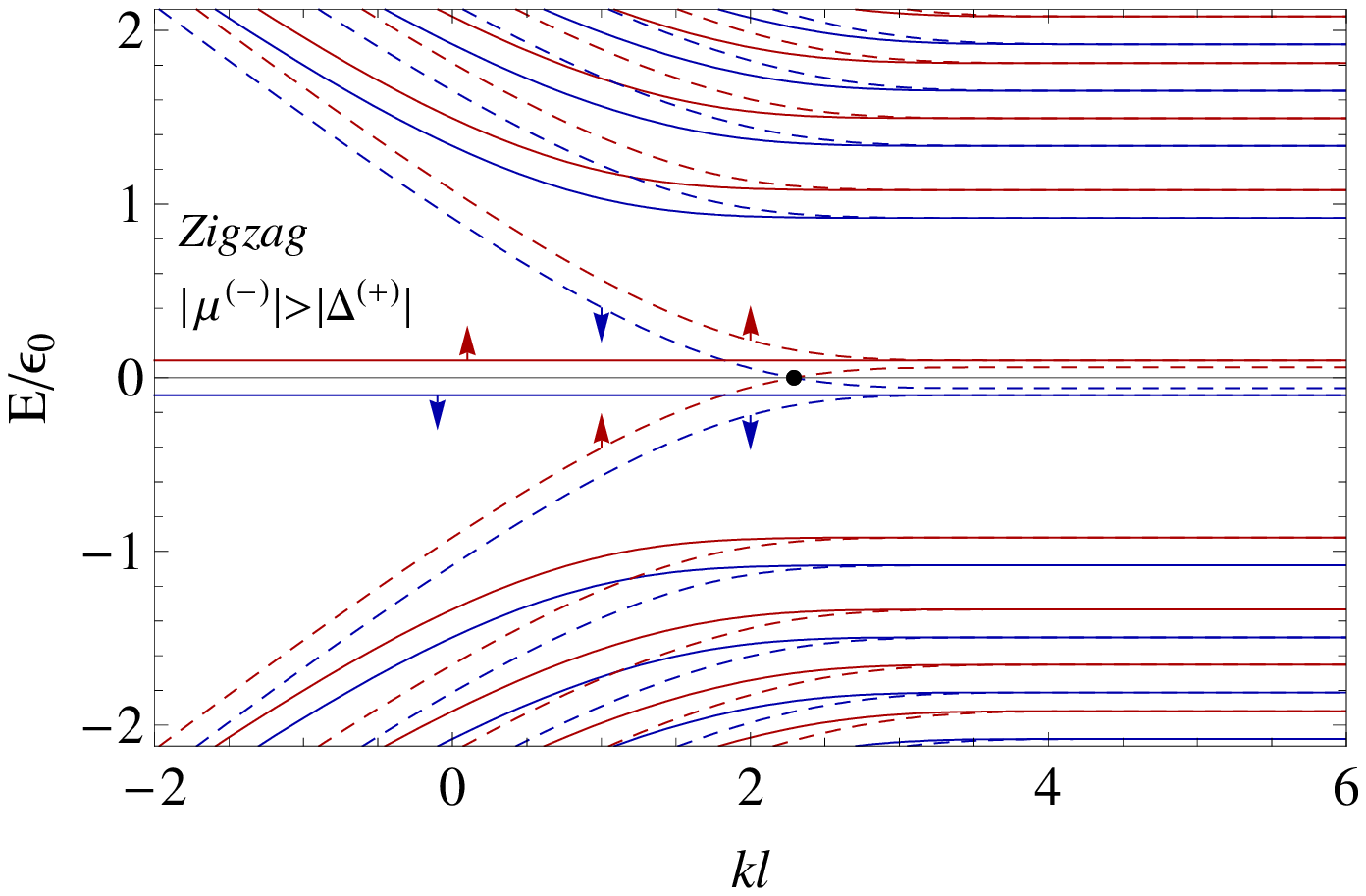}
\caption{Numerical results for the energy spectra of the first
few Landau levels near a zigzag edge of graphene in the case 
of nonzero spin splitting and nonzero {\em singlet} masses. 
The values of parameters are 
$\mu_{\pm}=\mp 0.02 \epsilon_0$ and $\Delta_{\pm}=\pm 0.08 \epsilon_0$
in the left panel, and 
$\mu_{\pm}=\mp 0.08 \epsilon_0$ and $\Delta_{\pm}=\pm 0.02 \epsilon_0$
in the right panel. (The subscript indices in $\mu_{\pm}$ and $\Delta_{\pm}$ 
denote the spin orientations.)
 In the first case $|\mu^{(-)}| < |\Delta^{(+)}|$ and there are
no gapless modes, in the second case $|\mu^{(-)}| > |\Delta^{(+)}|$ and 
gapless modes are present. Spin-up and spin-down states are denoted 
by red ($s=+$) and blue ($s=-$) color of the lines. In the lowest energy 
sublevels the spins are also marked by arrows. The spectra around 
$K_+$ ($K_-$) point are shown by solid (dashed) lines.}
\label{spectrum-zigzag2}
\end{center}
\end{figure}

Two specific examples of energy spectra, with and without gapless 
modes, are given in Fig.~\ref{spectrum-zigzag2}. In the left panel, 
the first few Landau levels in the case of a small spin gap, which is 
modeled by $\mu_{\pm}=\mp 0.02 \epsilon_0$ with the subscript index denoting 
the spin, and a larger {\em singlet} mass, which is given by $\Delta_{\pm}=\pm 0.08 
\epsilon_0$, are shown. Since $|\mu^{(-)}|< |\Delta^{(+)}|$, there are 
no gapless modes in this case. In the right panel of 
Fig.~\ref{spectrum-zigzag2}, the low-energy spectrum is shown 
for another choice of parameters, i.e., $\mu_{\pm}=\mp 0.08 \epsilon_0$ 
and $\Delta_{\pm}=\pm 0.02 \epsilon_0$, which satisfies the condition 
in Eq.~(\ref{cond-K-}). As expected, in this case there are 
gapless edge states in the spectrum. By taking into account the fact that the 
group velocities of gapless modes, $v_x=\partial E/\partial k|_{E=0}$, have 
opposite signs along the $x$-direction, the up- and down-spin states 
carry counter-propagating currents.\cite{Abanin2007PRL,Abanin2006PRL}
It is also curious to note that these gapless states are chiral since they 
belong to a single valley ($K_{-}$).
%%%%%%%%%%%%%%%%%%%%%%%%  Figure Zigzag-gapped-ungapped %%%%%%%%%%%%%
%%%%%%%%%%%%%%%%%%%%%%%%%%%%%%%%%%%%%%%%%%%%%%%%%%%%%%%%%%%%%%%%%%%%%

Before concluding this section, it might be appropriate to mention 
that the examples of spectra shown in Fig.~\ref{spectrum-zigzag2}
may have a direct application to the case of graphene in a strong
magnetic field. The corresponding choice of parameters with singlet, 
rather than triplet masses was taken in the same form as in the 
ground state around the neutral Dirac point, which was proposed in the 
dynamical model of Ref.~\onlinecite{GGM2007}. In fact, the spectra 
would look nearly the same also in the case of triplet masses, except 
perhaps for an overall shift of the dispersionless modes, which depend
not only on the absolute value but also on the sign of the mass terms. 

\section{Edge states for the armchair edge}
\label{sec:armchair}

\begin{figure}
\begin{center}
\includegraphics[width=.45\textwidth]{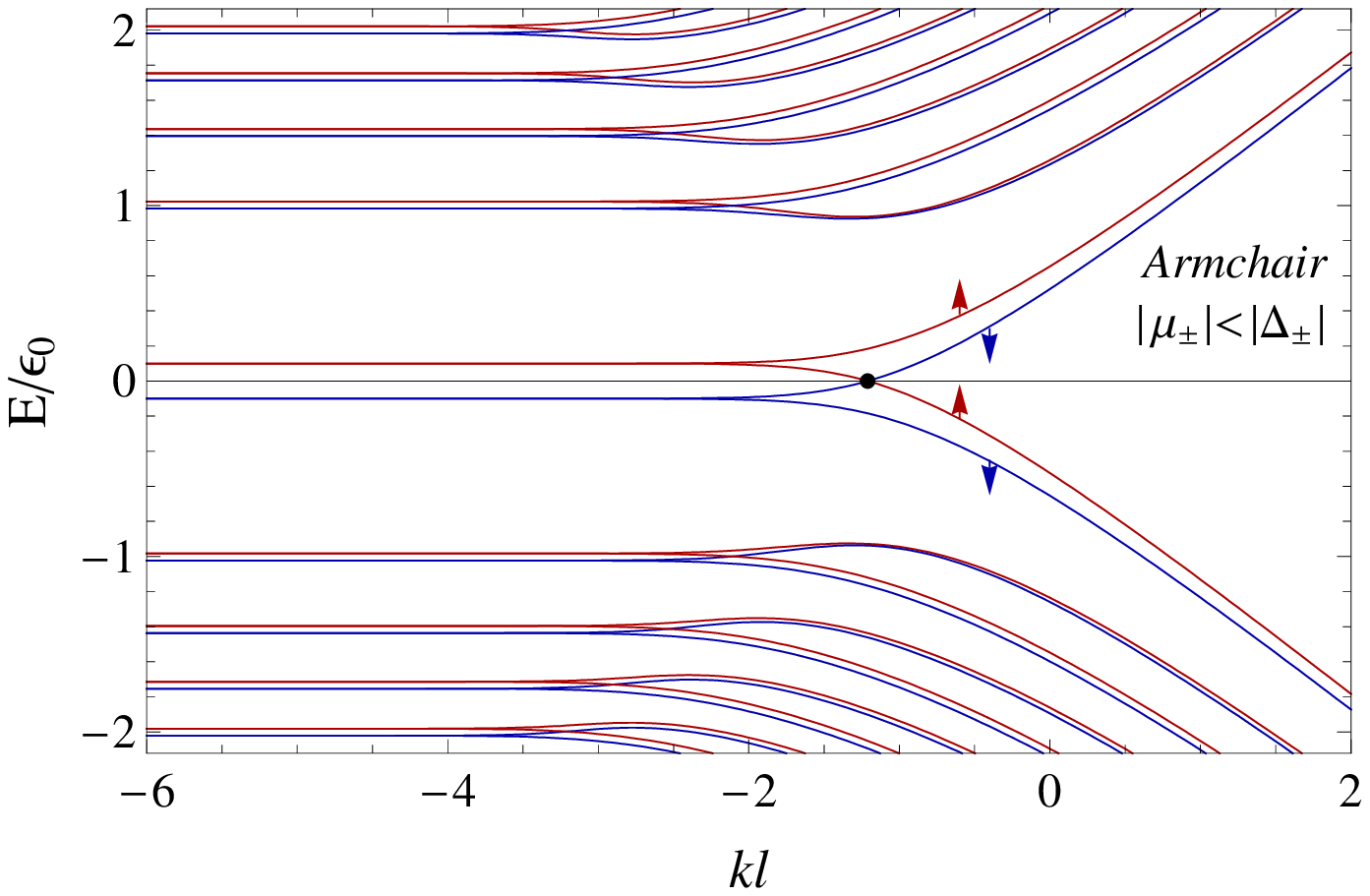}
\hspace{.04\textwidth}
\includegraphics[width=.45\textwidth]{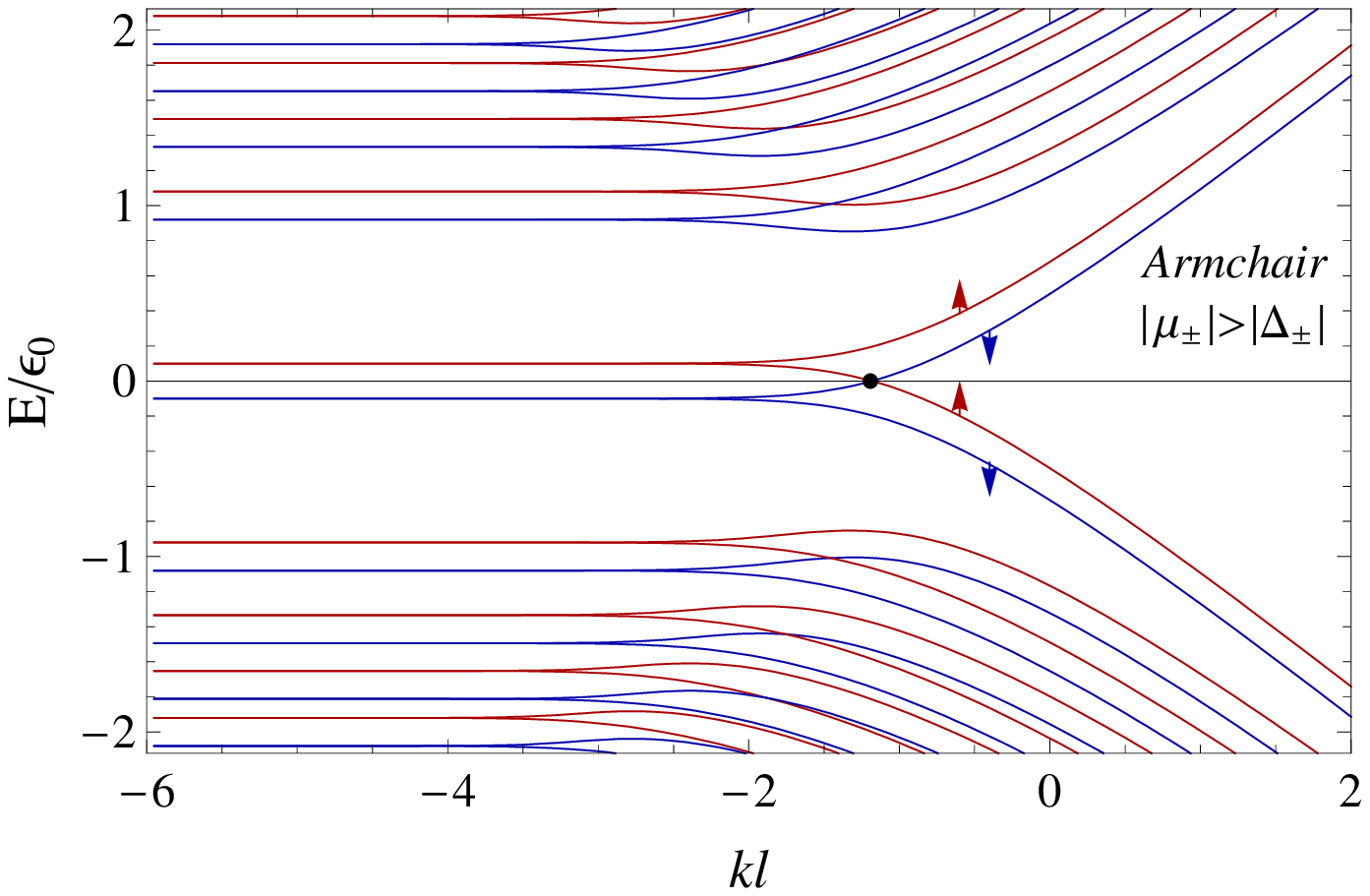}
\caption{Numerical results for the energy spectra of the first
few Landau levels near a armchair edge of graphene in the case 
of nonzero spin splitting and nonzero {\em singlet} masses. 
The values of parameters are 
$\mu_{\pm}=\mp 0.02 \epsilon_0$ and $\Delta_{\pm}=\pm 0.08 \epsilon_0$
in the left panel, and 
$\mu_{\pm}=\mp 0.08 \epsilon_0$ and $\Delta_{\pm}=\pm 0.02 \epsilon_0$
in the right panel. (The subscript indices in $\mu_{\pm}$ and $\Delta_{\pm}$ 
denote the spin orientations.)
In both cases, there are gapless modes in the spectrum. 
Spin-up and spin-down states are denoted by red ($s=+$) and blue ($s=-$) 
color of the lines. In the lowest energy sublevels the spins are also marked 
by arrows. }
\label{spectrum-armchair3}
\end{center}
\end{figure}

\begin{figure}
\begin{center}
\includegraphics[width=.45\textwidth]{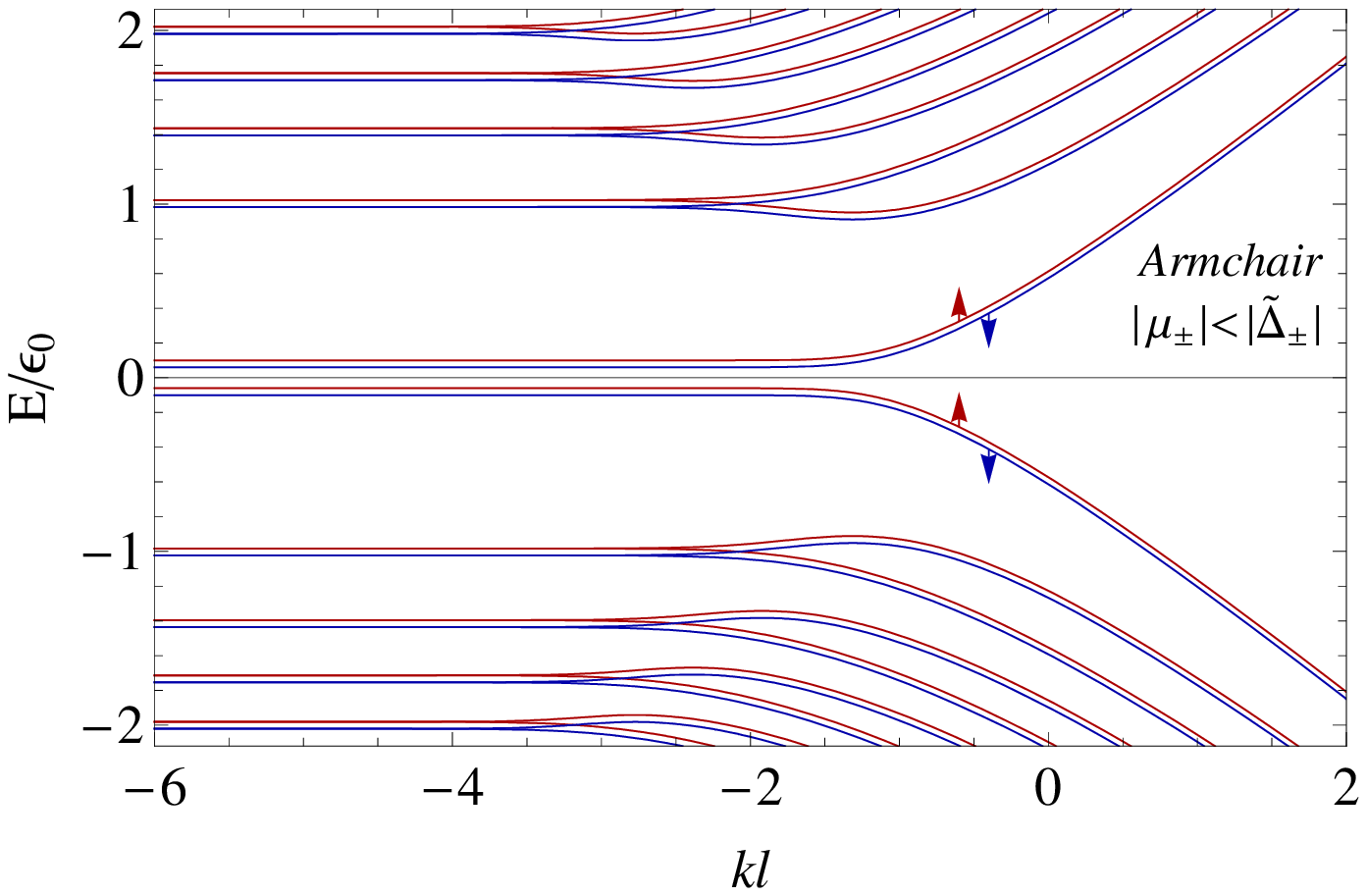}
\hspace{.04\textwidth}
\includegraphics[width=.45\textwidth]{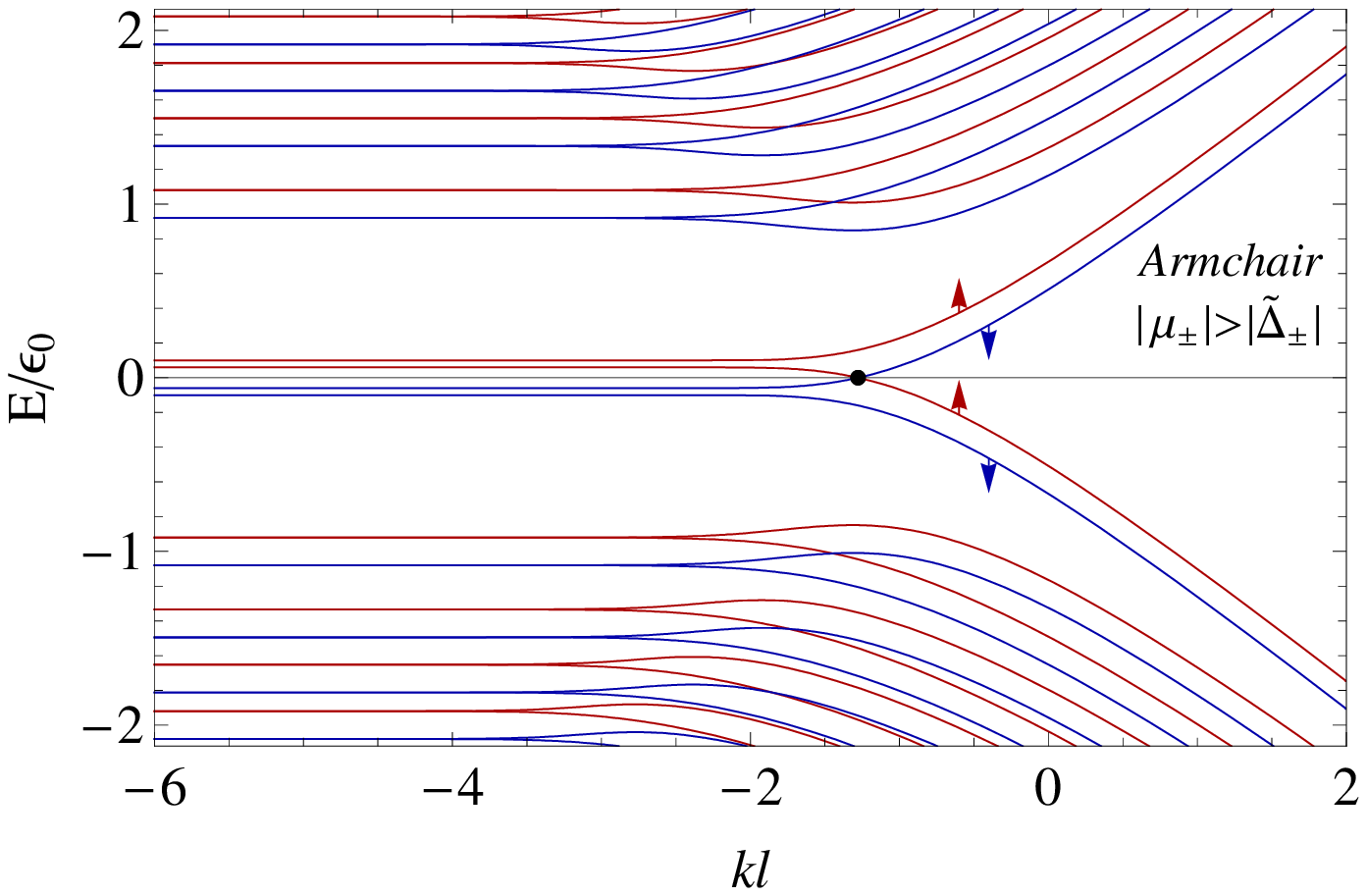}
\caption{Same as in Fig.~\ref{spectrum-armchair3}, but for the case of 
nonzero {\em triplet} masses. The values of parameters are 
$\mu_{\pm}=\mp 0.02 \epsilon_0$ and $\tilde{\Delta}_{\pm}=0.08 \epsilon_0$
in the left panel, and 
$\mu_{\pm}=\mp 0.08 \epsilon_0$ and $\tilde{\Delta}_{\pm}=0.02 \epsilon_0$
in the right panel. 
The existence of gapless modes depends on the relative magnitude of 
$|\mu_{\pm}|$  and $|\tilde{\Delta}_{\pm}|$.}
\label{spectrum-armchair2}
\end{center}
\end{figure}

In this section, we analyze the spectrum of edge modes in the case of an 
armchair edge. We take the armchair edge parallel to the $y$-direction, 
as shown in Fig.~\ref{illustration}. In this case, it is convenient to use 
a different Landau gauge with $(A_{x},A_{y})=(0,Bx)$. Accordingly, the 
solutions of Eq.~(\ref{differential-common}) are translation invariant along 
the $y$-direction,
\begin{equation}\label{y-armchair}
\begin{split}
\psi_{AK_{+}}(\mathbf{r},k)&=\frac{1}{\sqrt{2\pi
l}}e^{iky}u_{+}(x,k),\qquad \psi_{BK_{+}}=\frac{1}{\sqrt{2\pi
l}}e^{iky}v_{+}(x,k), \\
\psi_{AK_{-}}(\mathbf{r},k)&=\frac{1}{\sqrt{2\pi
l}}e^{iky}u_{-}(x,k), \qquad \psi_{BK_{-}}=\frac{1}{\sqrt{2\pi
l}}e^{iky}v_{-}(x,k).
\end{split}
\end{equation}
Then, the corresponding differential equations for functions
$u_{\pm}(x,k)$, which are rewritten in terms of the dimensionless variable
$\xi=x/l +kl$, coincide with Eq.~(\ref{differential-zigzag}). The
expressions for the eliminated components $v_{\pm}(\xi)$, 
however, slightly differ from Eq.~(\ref{v_pm-zigzag}), and 
are given by
\begin{equation}
\label{v_pm-armchair} 
v_{\pm}(\xi) =\mp
i\frac{\epsilon_0\left(\partial_{\xi}\mp\xi\right)u_{\pm}(\xi)}
{\sqrt{2}(E+\mu^{(\pm)} \mp\Delta^{(\mp)})}.
\end{equation}
We consider a graphene sheet in the half-plane $x>0$. Since the armchair 
edge has lattice sites of both  $A$ and $B$ types, the wave function should 
vanish at both these sites along the $x=0$ 
line,\cite{Falko2004JPCM,Brey2006PRB,Abanin2006PRL}
\begin{equation}
\begin{split}
&u_{+}(x=0)+u_{-}(x=0)=0, \\
&v_{+}(x=0)+v_{-}(x=0)=0.
\end{split}
\label{BC} 
\end{equation}
Note that armchair boundary conditions mix the chiralities associated with the $K_{+}$ and
$K_{-}$ valleys. The general solutions for the $u_{\pm}(\xi)$ functions have the same form 
as in Eqs.~(\ref{uPlus}) and (\ref{uMinus}), 
\begin{eqnarray}
\label{uPlusArm}
&&
u_{+}(\xi)= C_{1} \frac{E+\mu^{(+)}-\Delta^{(-)}}{\epsilon_0 }
U\left(\frac{1-2\lambda_{+}}{2},\sqrt{2}{\xi}\right)
+C_{2} V\left(\frac{1-2\lambda_{+}}{2},\sqrt{2}{\xi}\right),\\
\label{uMinusArm}
&&
u_{-}(\xi) = C_{3}U\left(-\frac{1+2\lambda_{-}}{2},\sqrt{2}{\xi}\right)
+C_{4} \frac{E+\mu^{(-)}+\Delta^{(+)}}{\epsilon_0}
V\left(-\frac{1+2\lambda_{-}}{2},\sqrt{2}{\xi}\right),
\end{eqnarray}
but with $\xi=x/l +kl$. By using the relations in Eqs.~(\ref{v_pm-armchair}) and 
(\ref{D-relations}), we also obtain the explicit expression for $v_{\pm}(\xi)$ 
functions,
\begin{eqnarray}
\label{vPlusArm}
&&
v_{+}({\xi})=iC_{1}
U\left(-\frac{1+2\lambda_{+}}{2},\sqrt{2}{\xi}\right)
+iC_{2}\frac{E+\mu^{(+)}+\Delta^{(-)}}{\epsilon_0 }
V\left(-\frac{1+2\lambda_{+}}{2},\sqrt{2}{\xi}\right),\\
\label{vMinusArm}
&&
v_{-}({\xi})=iC_{3} \frac{E+\mu^{(-)}-\Delta^{(+)}}{\epsilon_0}
U\left(\frac{1-2\lambda_{-}}{2},\sqrt{2}{\xi}\right)
+iC_{4}V\left(\frac{1-2\lambda_{-}}{2},\sqrt{2}{\xi}\right).
\end{eqnarray}
As in the zigzag case, here, normalizable wave functions are given 
in terms of only the $U(a,z)$-function, which falls off exponentially as
$z\to+\infty$, unlike the function $V(a,z)$, which grows exponentially
in both directions $z\to\pm\infty$. Therefore, we set $C_2=0$ and 
$C_4=0$. Then, the armchair boundary conditions [Eq.~(\ref{BC})] lead to 
the following system of equations:
\begin{equation}
\begin{split}
& C_{1} \frac{E+\mu^{(+)}-\Delta^{(-)}}{\epsilon_0} 
D_{\lambda_{+} -1}(\sqrt{2}kl)
+C_{3} D_{\lambda_{+} }(\sqrt{2}kl)=0,\\
& C_{1}D_{\lambda_{-} }(\sqrt{2}kl)
+C_{3}\frac{E+\mu^{(-)}-\Delta^{(+)}}{\epsilon_0} 
D_{\lambda_{-}  -1}(\sqrt{2}kl)=0,
\end{split}
\end{equation}
where again we used relation (\ref{UD-connection}) to rewrite the expression in 
terms of the parabolic cylinder function $D_{\nu}(z)$. This system has nontrivial 
solutions when the determinant of coefficient functions is zero, i.e.,
\begin{equation}
\frac{\left(E+\mu^{(+)}-\Delta^{(-)}\right)
\left(E+\mu^{(-)}-\Delta^{(+)}\right)}{\epsilon_0^2} 
D_{\lambda_{+} -1}\left(\sqrt{2}kl\right)
D_{\lambda_{-} -1}\left(\sqrt{2}kl\right)
-D_{\lambda_{+} }\left(\sqrt{2}kl\right)
D_{\lambda_{-} }\left(\sqrt{2}kl\right)=0.
\label{equation:armchair} 
\end{equation}
The numerical solutions to this equation for several representative choices 
of parameters are shown in Figs.~\ref{spectrum-armchair3} and \ref{spectrum-armchair2}.
%%%%%%%%%%%%%%%% Figure Figure Armchair-gapped-ungapped  %%%%%%%%%%%%%%%%%%%%%%%%%%%%
%%%%%%%%%%%%%%%%%%%%%%%%%%%%%%%%%%%%%%%%%%%%%%%%%%%%%%%%%%%%%%%%%%%%%
%%%%%%%%%%%%%%%% Figure Figure Armchair-gapped-ungapped  %%%%%%%%%%%%%%%%%%%%%%%%%%%%
%%%%%%%%%%%%%%%%%%%%%%%%%%%%%%%%%%%%%%%%%%%%%%%%%%%%%%%%%%%%%%%%%%%%%

The two cases with {\em singlet} masses are illustrated in Fig.~\ref{spectrum-armchair3}.
In the left panel, the first few Landau levels in the case of $\mu_{\pm}=\mp 0.02 
\epsilon_0$ and $\Delta_{\pm}=\pm 0.08 \epsilon_0$ are shown. In the right panel, 
instead, the corresponding values are $\mu_{\pm}=\mp 0.08 \epsilon_0$ and 
$\Delta_{\pm}=\pm 0.02 \epsilon_0$. Note that here $\tilde{\mu}_{\pm}=
\tilde{\Delta}_{\pm}=0$. (Here, we restored the subscript indices which denote 
the quasiparticle spin orientations.) As we can see, in both cases the spectra contain 
gapless edge states. This is in strong contrast to the zigzag edge case. Indeed, 
for the armchair edge, gapless modes exist irrespective of the actual relation between 
the values of the {\em singlet} masses and spin splitting gaps. In part, this property 
could be understood from the topology of the spectra around the edge and the fact that the 
singlet mass does not break the $SU(2)_{s}$ valley symmetry. The double degenerate 
sublevels with a given spin, which should exist in the bulk because of the $SU(2)_{s}$ 
symmetry, repel in opposite directions near the edge. Then, 
gapless modes become almost inevitable at the edge. 

We note that the gapless edge states in Fig.~\ref{spectrum-armchair3} consist of a pair
of opposite spin states, carrying counter-propagating currents along the edge. This 
is qualitatively the same situation as found in Ref.~\onlinecite{Abanin2006PRL}. 
Interestingly, though, if the values of singlet masses $\Delta_{+}$ and $\Delta_{-}$
had the same signs, the opposite spin states would carry currents in the {\em same} 
direction along the edge. The observational implications of this fact could be quite 
unusual. It is not clear, however, if such a state can be realized since the dynamical 
model of Ref.~\onlinecite{GGM2007} indicates that singlet masses $\Delta_{+}$ 
and $\Delta_{-}$ should have opposite signs in the ground state.

The two cases with {\em triplet} masses are illustrated in Fig.~\ref{spectrum-armchair2}.
The values of the parameters in these cases are (i) $\mu_{\pm}=\mp 0.02 
\epsilon_0$ and $\tilde{\Delta}_{\pm}=0.08 \epsilon_0$ (left panel in 
Fig.~\ref{spectrum-armchair2}) and (ii)  $\mu_{\pm}=\mp 0.08 \epsilon_0$ 
and $\tilde{\Delta}_{\pm}=0.02 \epsilon_0$ (right panel in 
Fig.~\ref{spectrum-armchair2}). These energy spectra resemble the spectra
for the zigzag edge, studied in Sec.~\ref{sec:zigzag}. There are no gapless edge 
states when the mass is larger than the spin splitting, and there are such states 
when the mass is smaller than the spin splitting.  

%%%%%%%%%%%%%%%%%%%%%%%%%%%%  Figure Armchair  %%%%%%%%%%%%%%%%%%%%%%%%%%%%
\begin{figure}
\begin{center}
\includegraphics[width=.48\textwidth]{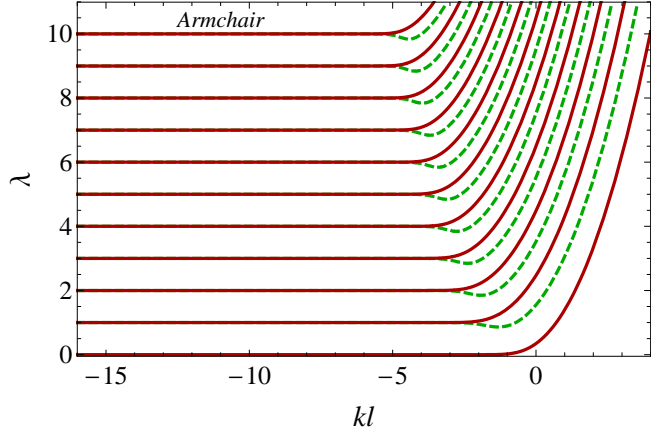}
\caption{Numerical solutions of Eq.~(\ref{equation:arm-triplet}) for the dimensionless 
parameter $\lambda$ in the case of an armchair boundary. This is valid for a general 
choice of $\tilde{\Delta}$ and $\mu$, but only if $\tilde{\mu}$ and $\Delta$ vanish.}
\label{spectrum-armchair1}
\end{center}
\end{figure}
%%%%%%%%%%%%%%%%%%%%%%%%%%%%%%%%%%%%%%%%%%%%%%%%%%%%%%%%%%%%%%%%%%%%%

In fact, in the case of the triplet mass $\tilde{\Delta}$ and a nonzero $\mu$
(but vanishing $\tilde{\mu}$ and $\Delta$), we can study the energy spectra
around the armchair edge in a general case, just like we did for the zigzag 
edge. In this particular case, the spectral equation (\ref{equation:armchair}) 
takes the following simple form:
\begin{equation} 
\lambda D^{2}_{\lambda-1}\left(\sqrt{2}kl\right)-D^{2}_{\lambda}\left(\sqrt{2}kl\right)=0,
\label{equation:arm-triplet} 
\end{equation}
where $\lambda=[(E+\mu)^2-\tilde{\Delta}^2]/\epsilon_0^2$. By expressing 
$\lambda$ in terms of squares of parabolic cylinder functions from 
Eq.~(\ref{equation:arm-triplet}), we see that solutions to this equation 
exist only with $\lambda\geq 0$. Therefore, the energy spectrum takes the 
following form:  
\begin{equation}
E_{n}(k) = -\mu \pm \sqrt{ \lambda (kl, n)\epsilon_0^2 +\tilde{\Delta}^2},
\quad\mbox{where}\quad
\lambda(kl, n)\geq 0,
\label{spectr_arm-triplet}
\end{equation}
where $n=0,1,2,\ldots$. Additionally, one can show that $\lambda(kl, n)\simeq n$ 
when $|kl|\gg 1$ and $k$ is negative. Our numerical results for $\lambda$ as 
a function of $kl$ are presented in Fig.~\ref{spectrum-armchair1}. By combining
the numerical information with the general expression for the energy 
(\ref{spectr_arm-triplet}), we see that the necessary and sufficient condition 
for having gapless modes is $|\mu|\geq |\tilde{\Delta}|$.

\section{Discussion}
\label{sec:concl}

In this paper, we studied the spectra of edge states in graphene on 
a half-plane with zigzag and armchair boundary conditions, and  
derived the conditions for the existence of the gapless edge states 
for various types of masses and chemical potentials that could be 
spontaneously generated in QHE, e.g., at $\nu=0$ and $\nu=\pm 1$
plateaus. 

Our analysis of singlet and triplet Dirac masses [with respect to 
the valley symmetry group $SU(2)_s$] shows that spectral 
properties of zigzag and armchair edges are affected by  
(i) the relative magnitude of the masses and chemical potentials,
and (ii) the types of masses. In particular, we found the criteria 
for the existence of gapless edge states in the spectra. These 
can be summarized as follows.

\begin{itemize}
\item[(i)] {\em Zigzag edge}:
the necessary and sufficient condition to have a gapless state is 
that at least one of the following inequalities is satisfied:
\begin{eqnarray}
|\mu_{s}^{(+)}| &\geq & \sqrt{\epsilon_0^{2} +\left(\Delta_{s}^{(-)}\right)^{2}},
\label{cond111}\\
|\mu_{s}^{(-)}| &\geq & |\Delta_{s}^{(+)}|. 
\label{cond222}
\end{eqnarray}

\item[(ii)] {\em Armchair edge}:
	\begin{itemize}
	\item[(a)] gapless edge states exist always when there are {\em singlet} 
					Dirac masses, irrespective of the actual relation between the 
					values of the masses and the chemical potentials;
	\item[(b)] in the case of {\em triplet} Dirac masses, gapless edge states
					exist if $|\mu_{\pm}|>|\tilde{\Delta}_{\pm}|$, and do not exist 
					otherwise.
	\end{itemize}
\end{itemize}
These conditions are consistent with the two limiting cases,
analyzed in Ref.~\onlinecite{Abanin2006PRL}. Also, the results in this paper 
extend our previous findings in the case of a graphene ribbon with zigzag 
edges.\cite{Gusynin2008} The situation on a half-plane with a zigzag edge 
is essentially the same one, modulo the fact that there is one edge instead of 
two.

The above criteria are derived for ideal, smooth edges and for a 
perfect graphene layer without disorder. In reality, the available graphene 
samples are disordered. Because of the geometrical roughness and 
impurities, they do not have perfect zigzag or armchair edges either. 
Then, the corresponding boundary conditions for the graphene 
wave functions may be different from those used in the current 
study.\cite{roughedges} Additionally, the bonds of 
the carbon atoms at the edges can be saturated by foreign atoms 
modifying even perfectly smooth and regular edges.\cite{Dutta} 
Therefore, it is of great importance to study the effects of various 
types of disorder in graphene. This is, however, 
beyond the scope of the present paper. Here we limit our study to an 
idealized model in order to provide a clean benchmark calculation 
before a more detailed investigation of disorder is undertaken. 
By taking into account a considerable improvement in sample quality 
seen in graphene suspended above a graphite substrate\cite{Andrei} 
or above a Si/SiO$_2$ gate electrode\cite{Bolotin}, it is possible 
that the clean limit already provides a reasonable qualitative 
description of edge states. Additionally, because of the special 
nature of the LLL, the role of some types of disorder may be strongly 
suppressed.\cite{Giesbers2007PRL} For example, the effect of 
the randomness in the bond couplings and in the on-site potential 
caused by short range interactions is 
studied in Ref.~\onlinecite{Koshino2007PRB}. It is shown that the 
degeneracy of $K_{\pm}$ points is not lifted by the on-site disorder, 
but can be removed by the randomness in the bond couplings.

The results here are of interest in connection with the interpretation of the 
$\nu=0$ Hall plateau. Indeed, the gapless edge states should play an important role 
in the charge transport of graphene in a strong magnetic field. Their presence 
is expected to make graphene a so-called quantum Hall metal, while their 
absence should make it an insulator.\cite{Abanin2007PRL,Abanin2006PRL}  
The actual temperature dependence of the longitudinal resistivity at the 
$\nu=0$ plateau in Refs.~\onlinecite{Zhang2006PRL} and \onlinecite{Abanin2007PRL} 
is consistent with the metal type. This conclusion may be disputed in view 
of the recent data from Ref.~\onlinecite{Ong2007} that reveal a clear plateau
at $\nu=0$, but the temperature dependence of the diagonal component 
of the resistivity signals a crossover to an insulating state in high fields. 
The latter observations do not seem to support the existence of gapless 
edge states. 

Our analysis in this paper as well as in Ref.~\onlinecite{Gusynin2008} 
suggests that the conditions for the existence and absence of gapless
edge states sensitively depend on the values of QHF and MC order 
parameters that characterize the nature of the corresponding QH state. 
Moreover, the microscopic analysis of Ref.~\onlinecite{GGM2007} indicates 
that the order parameters of both types necessarily coexist. Therefore, the 
dynamics is very likely to be rich and full of surprises. The situation with 
the edge states is probably just one of such surprises.

\acknowledgments

The authors acknowledge useful discussions with E.V.~Gorbar, 
H.~Fertig, I.F.~Herbut, M.I.~Katsnelson, L.~Levitov, and B.I.~Shklovskii.
V.P.G. and S.G.S. thank A.K.~Geim for the discussion of the 
experimental data that indicate the existence of the gapless edge
states in graphene. The work of V.P.G. was supported by the 
SCOPES Project No. IB 7320-110848 of the NSF-CH, Grant No. 10/07-N 
``Nanostructure systems, nanomaterials, nanotechnologies," 
and the Program of Fundamental Research of the Physics and 
Astronomy Division of the National Academy of Sciences of Ukraine. 
The work of V.A.M. was supported by the Natural Sciences and 
Engineering Research Council of Canada.


\begin{thebibliography}{99}
\bibitem{foot1} In both cases, the $SU(4)$ symmetry is associated with
spin-pseudospin degrees of freedom, but in graphene, pseudospin
is related to the sublattice index, in bilayer quantum Hall systems, pseudospin 
is related to the layer index.
\bibitem{foot2} Strictly speaking, in order to preserve
the $SU(2)_{s}$ valley symmetry of the model, all three components of the
triplet should be included in the Lagrangian density on equal footing. The
ground state, however, will correspond to a specific ``vacuum alignment",
e.g., characterized by a nonvanishing vacuum expectation value of the operator
$\bar{\Psi}_{s} \gamma^3\Psi_{s}$.
\end{thebibliography}
\end{document}